\documentclass[]{spie}

\usepackage[]{graphicx}

\title{\Large \bf Spin Networks and Anyonic Topological
Computing II }

\author{Louis H. Kauffman\supit{a} and Samuel J. Lomonaco Jr.\supit{b}
\skiplinehalf
\supit{a} Department of Mathematics, Statistics and Computer Science  
(m/c 249), 851 South Morgan Street, University of Illinois at Chicago,
Chicago, Illinois 60607-7045, USA \\
\supit{b} Department of Computer Science and Electrical Engineering, University of
Maryland Baltimore County, 1000 Hilltop Circle, Baltimore, MD 21250, USA}
 
\authorinfo{Further author information: L.H.K.  E-mail: kauffman@uic.edu,
S.J.L. Jr.: E-mail: lomonaco@umbc.edu}
 
\begin{document} 

  \maketitle 

\begin{abstract} 
We review the $q$-deformed spin network approach to Topological Quantum Field Theory and apply these methods to
produce unitary representations of the braid groups that are dense in the unitary groups. The simplest case of these models is
the Fibonacci model, itself universal for quantum computation. We here formulate these braid group 
representations in a form suitable for computation and algebraic work.
\end{abstract}


\section{INTRODUCTION}

This paper describes the background for topological quantum computing in terms of Temperely -- Lieb Recoupling Theory and gives an explicit description of
the resulting unitary representations of the Artin braid group, including the Fibonacci model as the simplest case.
\bigbreak

We use a recoupling theory that generalizes standard angular momentum recoupling theory, generalizes the Penrose theory of spin networks and is inherently
topological.  Temperely -- Lieb Recoupling Theory is based on the bracket polynomial model for the Jones polynomial. It is built in terms of diagrammatic
combinatorial topology. The same structure can be explained in terms of the $SU(2)_{q}$ quantum group, and has relationships with functional integration and
Witten's approach to topological quantum field theory. Nevertheless, the approach given here will be unrelentingly elementary. Elementary, does not necessarily
mean simple. In this case an architecture is built from simple beginnings and this archictecture and its recoupling language can be applied to many things
including: colored Jones polynomials, Witten--Reshetikhin--Turaev invariants of three manifolds, topological quantum field theory and quantum computing.
\bigbreak

The contents of this paper are based upon the work in \cite{KLSpie} and we shall refer to results from that paper.
\bigbreak

In quantum computing, the application is most interesting because the recoupling theory
yields representations of the Artin Braid group into unitary groups
$U(n)$. These represententations are {\em dense} in the unitary group, and can be used to model quantum computation
universally in terms of representations of the braid group. Hence the term: topological quantum computation.
\bigbreak

In this paper, we outline the basics of the Temperely -- Lieb Recoupling Theory, and show explicitly how unitary representations of the braid group arise from it.
We will return to this subject in more detail in subsequent papers. In particular, we do not describe the context of anyonic models for quantum computation
in this paper. Rather, we concentrate here on showing how naturally unitary representations of the braid group arise in the context of the Temperely -- Lieb
Theory. For the reader interested in the relevant background in anyonic topological quantum computing we recommend the following references 
\{ \cite{F,FR98,FLZ,F5,F6,Kitaev,MR,Preskill,Wilczek} \}.
\bigbreak

Here is a very condensed presentation of how unitary representations of the
braid group are constructed via topological quantum field theoretic methods.
For simplicity assmue that one has a single (mathematical) particle with label $P$
that can interact with itself to produce either itself labeled $P,$ or itself
with the null label $*.$ When $*$ interacts with $P$ the result is always $%
P. $ When $*$ interacts with $*$ the result is always $*.$ One considers
process spaces where a row of particles labeled $P$ can successively
interact, subject to the restriction that the end result is $P.$ For example
the space $V[(ab)c]$ denotes the space of interactions of three particles
labeled $P.$ The particles are placed in the positions $a,b,c.$  Thus we
begin with $(PP)P.$ In a typical sequence of interactions, the first two $P$%
's interact to produce a $*,$ and the $*$ interacts with $P$ to produce $P.$ 
\[
(PP)P \longrightarrow (*)P \longrightarrow P. 
\]
\noindent In another possibility, the first two $P$'s interact to produce a $%
P,$ and the $P$ interacts with $P$ to produce $P.$ 
\[
(PP)P \longrightarrow (P)P \longrightarrow P. 
\]
It follows from this analysis that the space of linear combinations of
processes $V[(ab)c]$ is two dimensional. The two processes we have just
described can be taken to be the the qubit basis for this space. One obtains
a representation of the three strand Artin braid group on $V[(ab)c]$ by
assigning appropriate phase changes to each of the generating processes. One
can think of these phases as corresponding to the interchange of the
particles labeled $a$ and $b$ in the association $(ab)c.$ The other operator
for this representation corresponds to the interchange of $b$ and $c.$ This
interchange is accomplished by a {\it unitary change of basis mapping} 
\[
F:V[(ab)c] \longrightarrow V[a(bc)]. 
\]
\noindent If 
\[
A:V[(ab)c] \longrightarrow V[(ba)c:d] 
\]
is the first braiding operator (corresponding to an interchange of the first
two particles in the association) then the second operator 
\[
B:V[(ab)c] \longrightarrow V[(ac)b] 
\]
is accomplished via the formula $B = F^{-1}AF$ where the $A$ in this formula
acts in the second vector space $V[a(bc)]$ to apply the phases for the
interchange of $b$ and $c.$ \bigbreak

In this scheme, vector spaces corresponding to associated strings of
particle interactions are interrelated by {\it recoupling transformations}
that generalize the mapping $F$ indicated above. A full representation of
the Artin braid group on each space is defined in terms of the local
intechange phase gates and the recoupling transfomations. These gates and
transformations have to satisfy a number of identities in order to produce a
well-defined representation of the braid group. These identities were
discovered originally in relation to topological quantum field theory. In
our approach the structure of phase gates and recoupling
transformations arise naturally from the structure of the bracket model for
the Jones polynomial \cite{JO}. Thus we obtain a knot-theoretic basis for topological
quantum computing. \bigbreak

\section {Spin Networks and Temperley -- Lieb Recoupling Theory}
In this section we discuss a combinatorial construction for spin networks that generalizes the original construction of Roger Penrose \cite{Penrose}.
The result of this generalization is a structure that satisfies all the properties of a graphical $TQFT$ as described in our paper on braiding and universal
quantum gates
\cite{UG}, and  specializes to classical angular momentum recoupling theory in the limit of its basic variable. The construction is based on the properties of 
the bracket polynomial \cite{KA87}. A complete description of this theory can be found in the book ``Temperley -- Lieb
Recoupling Theory and Invariants of Three-Manifolds" by Kauffman and Lins \cite{KL}.  
\bigbreak

The ``$q$-deformed" spin networks that we construct here are based on the bracket polynomial relation. View Figure 1 and Figure 2.
\bigbreak

\begin{figure}
     \begin{center}
     \begin{tabular}{c}
     \includegraphics[height=7cm]{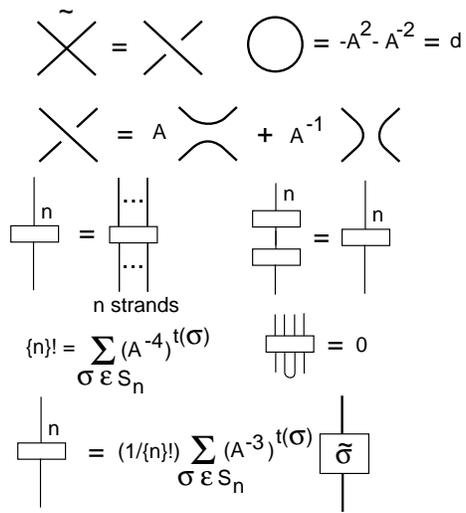}
     \end{tabular}
     \end{center}
     \caption{\bf Basic Projectors}
     \end{figure} 
     \bigbreak

\begin{figure}
     \begin{center}
     \begin{tabular}{c}
     \includegraphics[height=7cm]{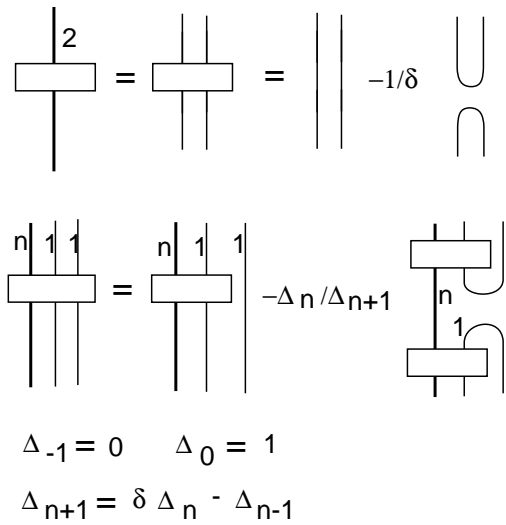}
     \end{tabular}
     \end{center}
     \caption{\bf Two Strand Projector}
     \end{figure} 
     \bigbreak

\begin{figure}
     \begin{center}
     \begin{tabular}{c}
     \includegraphics[height=4cm]{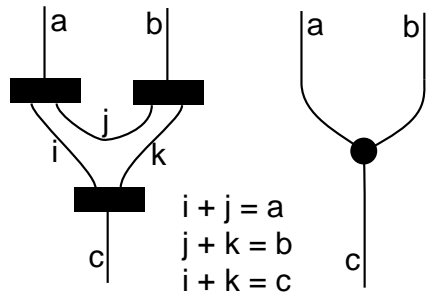}
     \end{tabular}
     \end{center}
     \caption{\bf Trivalent Vertex}
     \end{figure} 
     \bigbreak

In Figure 1 we indicate how the basic projector (symmetrizer, Jones-Wenzl projector) is constructed on the basis of the
bracket polynomial expansion \cite{KA87}. In this technology, a symmetrizer is a sum of tangles on $n$ strands (for a chosen integer $n$). The tangles are made by
summing over braid lifts of  permutations in the symmetric group on $n$ letters, as indicated in Figure 1. Each elementary braid is then expanded by the
bracket polynomial  relation, as indicated in Figure 1, so that the resulting sum  consists of flat tangles without any crossings (these can be viewed as
elements in the Temperley -- Lieb algebra). The projectors have the property that the concatenation of a projector with itself is just that projector, and
if you tie two lines on the top or the bottom of a projector together, then the evaluation is zero. This general definition of projectors is very useful for 
this theory. The two-strand projector is shown in Figure 2. Here the formula for that projector 
is particularly simple. It is the sum of two parallel arcs and two turn-around arcs (with coefficient $-1/d,$  with $d = -A^{2} - A^{-2}$ is the loop
value for the bracket polynomial. Figure 2 also shows the recursion formula for the general projector. This recursion formula is due to Jones and Wenzl and
the projector in this form, developed as a sum in the Temperley -- Lieb algebra (see Section 5 of this paper), is usually known as the {\em Jones--Wenzl
projector}.
\bigbreak

The projectors are combinatorial analogs of irreducible representations of a group (the original spin nets were based
on $SU(2)$ and these deformed nets are based on the quantum group corresponding to SU(2)). As such the reader can think of them as ``particles". The
interactions of these particles are governed by how they can be tied together into three-vertices. See Figure 3.
In Figure 3 we show how to tie three projectors, of $a,b,c$ strands respectively, together to form a three-vertex. In order to accomplish this 
interaction, we must share lines between them as shown in that Figure so that there are non-negative integers $i,j,k$ so that
$a = i + j, b = j + k, c = i + k.$ This is equivalent to the condition that $a + b + c$ is even and that the sum of any two of $a,b,c$ is 
greater than or equal to the third. For example $a + b \ge c.$ One can think of the vertex as a possible particle interaction where
$[a]$ and $[b]$ interact to produce $[c].$ That is, any two of the legs of the vertex can be regarded as interacting to produce the third leg.
\bigbreak

There is a basic orthogonality of three vertices as shown in Figure 4. Here if we tie two three-vertices together
so that they form a ``bubble" in the middle, then the resulting network with labels $a$ and $b$ on its free ends
is a multiple of an $a$-line (meaning a line with an $a$-projector on it) or zero (if $a$ is not equal to $b$).
The multiple is compatible with the results of closing the diagram in the equation of Figure 4 so the the two free
ends are identified with one another. On closure, as shown in the Figure, the left hand side of the equation becomes
a Theta graph and the right hand side becomes a multiple of a ``delta" where $\Delta_{a}$ denotes the bracket 
polynomial evaluation of the $a$-strand loop with a projector on it. The $\Theta(a,b,c)$ denotes the bracket 
evaluation of a theta graph made from three trivalent vertices and labeled with $a, b, c$ on its edges.

\begin{figure}
     \begin{center}
     \begin{tabular}{c}
     \includegraphics[height=7cm]{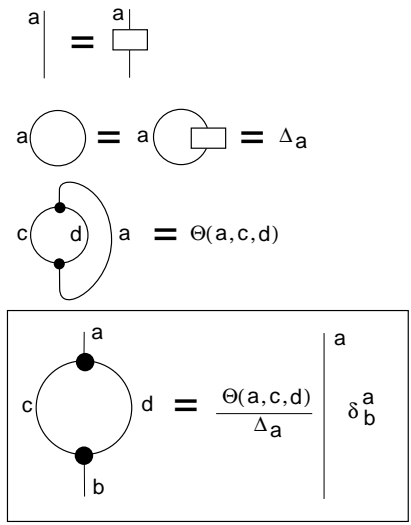}
     \end{tabular}
     \end{center}
     \caption{\bf Orthogonality of Trivalent Vertices}
     \end{figure} 
     \bigbreak

There is a recoupling formula in this theory in the form shown in Figure 5.
Here there are ``$6$-j symbols", recoupling coefficients that can be expressed, as shown in 
Figure 7, in terms of tetrahedral graph evaluations and theta graph evaluations. The tetrahedral graph is shown in 
Figure 6. One derives the formulas for 
these coefficients directly from the orthogonality relations for the trivalent vertices by 
closing the left hand side of the recoupling formula and using orthogonality to evaluate the right hand side.
This is illustrated in Figure 7.

\begin{figure}
     \begin{center}
     \begin{tabular}{c}
     \includegraphics[height=3cm]{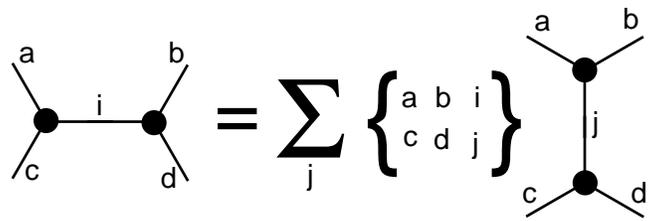}
     \end{tabular}
     \end{center}
     \caption{\bf Recoupling Formula }
     \end{figure} 
     \bigbreak

\begin{figure}
     \begin{center}
     \begin{tabular}{c}
     \includegraphics[height=2cm]{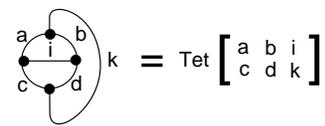}
     \end{tabular}
     \end{center}
     \caption{\bf Tetrahedron Network}
     \end{figure} 
     \bigbreak

\begin{figure}
     \begin{center}
     \begin{tabular}{c}
     \includegraphics[height=8cm]{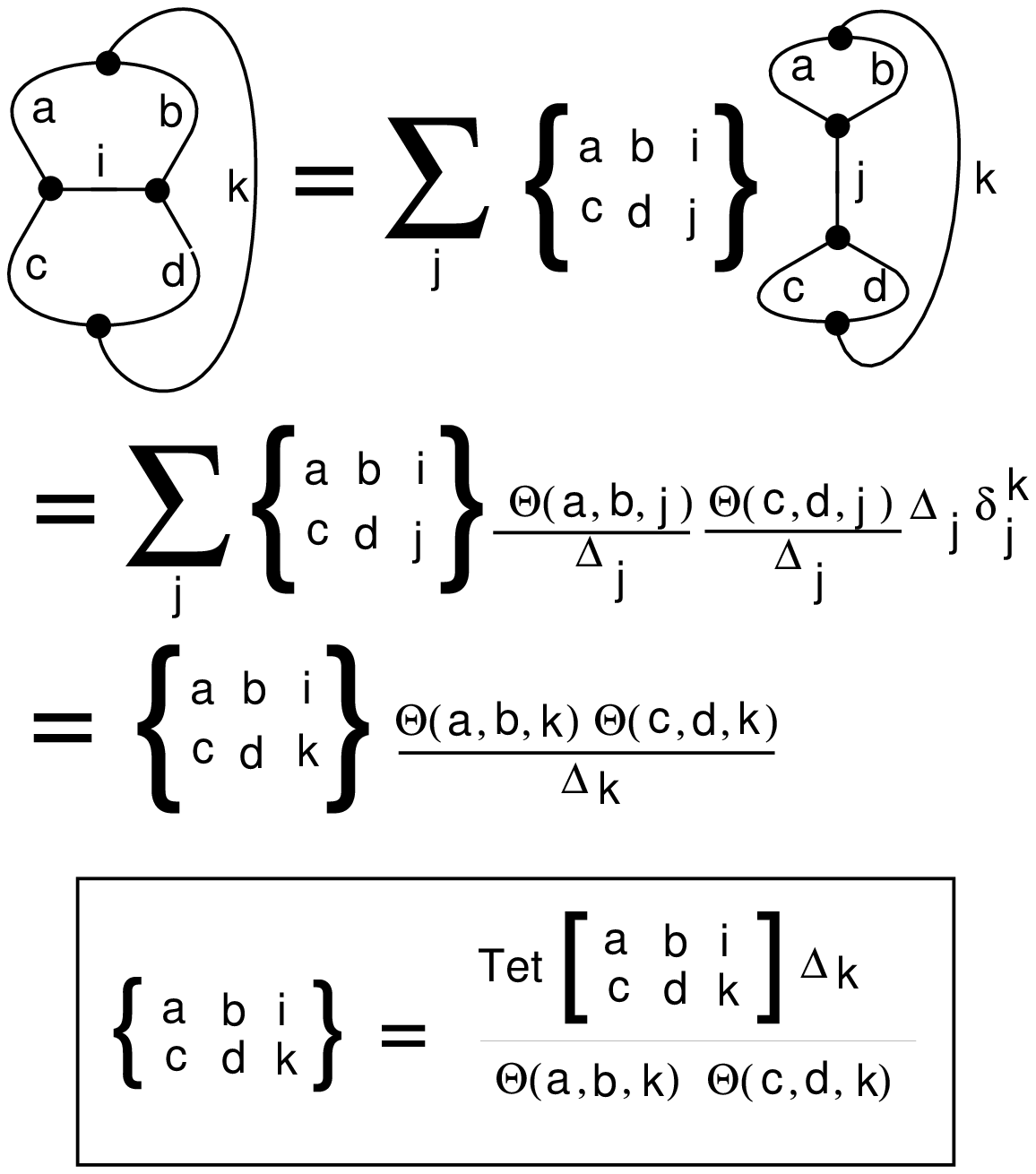}
     \end{tabular}
     \end{center}
     \caption{\bf Tetrahedron Formula for Recoupling Coefficients}
     \end{figure} 
     \bigbreak

Finally, there is the braiding relation, as illustrated in Figure 8.

\begin{figure}
     \begin{center}
     \begin{tabular}{c}
     \includegraphics[height=4cm]{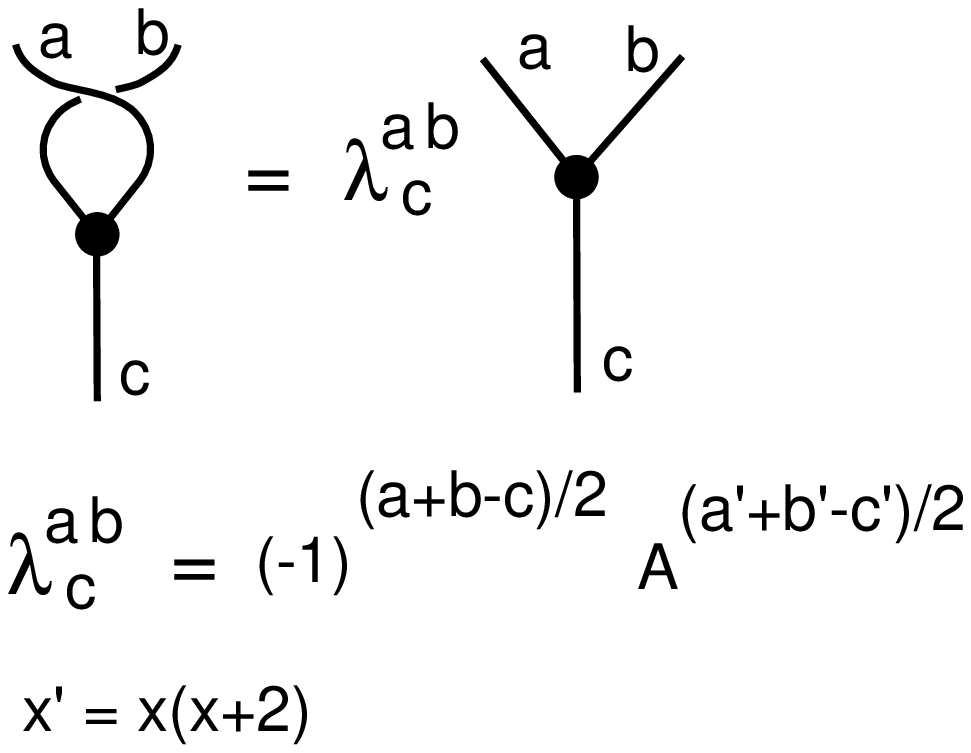}
     \end{tabular}
     \end{center}
     \caption{\bf LocalBraidingFormula}
     \end{figure} 
     \bigbreak

With the braiding relation in place, this $q$-deformed spin network theory satisfies the pentagon, hexagon and braiding naturality identities
needed for a topological quantum field theory. All these identities follow naturally from the basic underlying topological construction of the 
bracket polynomial. One can apply the theory to many different situations.

\subsection{Evaluations}
In this section we discuss the structure of the evaluations for $\Delta_{n}$ and the theta and tetrahedral networks. We refer to 
\cite{KL} for the details behind these formulas. Recall that $\Delta_{n}$ is the bracket evaluation of the closure of the $n$-strand
projector, as illustrated in Figure 4. For the bracket variable $A,$ one finds that 
$$\Delta_{n} = (-1)^{n}\frac{A^{2n+2} - A^{-2n-2}}{A^{2} - A^{-2}}.$$
One sometimes writes the {\it quantum integer}
$$[n] = (-1)^{n-1}\Delta_{n-1} = \frac{A^{2n} - A^{-2n}}{A^{2} - A^{-2}}.$$
If $$A=e^{i\pi/2r}$$ where $r$ is a positive integer, then 
$$\Delta_{n} = (-1)^{n}\frac{sin((n+1)\pi/r)}{sin(\pi/r)}.$$
Here the corresponding quantum integer is
$$[n] = \frac{sin(n\pi/r)}{sin(\pi/r)}.$$
Note that $[n+1]$ is a positive real number for $n=0,1,2,...r-2$ and that $[r-1]=0.$
\bigbreak

The evaluation of the theta net is expressed in terms of quantum integers by the formula
$$\Theta(a,b,c) = (-1)^{m + n + p}\frac{[m+n+p+1]![n]![m]![p]!}{[m+n]![n+p]![p+m]!}$$
where $$a=m+p, b=m+n, c=n+p.$$ Note that $$(a+b+c)/2 = m + n + p.$$
\bigbreak

When $A=e^{i\pi/2r},$ the recoupling theory becomes finite with the restriction that only three-vertices
(labeled with $a,b,c$) are {\it admissible} when $a + b +c \le 2r-4.$ All the summations in the 
formulas for recoupling are restricted to admissible triples of this form.
\bigbreak

\subsection{Symmetry and Unitarity}
The formula for the recoupling coefficients given in Figure 7 has less symmetry than is actually inherent in the structure of the situation.
By multiplying all the vertices by an appropriate factor, we can reconfigure the formulas in this theory so that the revised recoupling transformation is
orthogonal, in the sense that its transpose is equal to its inverse. This is a very useful fact. It means that when the resulting matrices are real, then
the recoupling transformations are unitary. 
\bigbreak

Figure 9 illustrates this modification of the three-vertex. Let $Vert[a,b,c]$ denote the original $3$-vertex of the Temperley -- Lieb recoupling theory.
Let $ModVert[a,b,c]$ denote the modified vertex. Then we have the formula
$$ModVert[a,b,c] = \frac{\sqrt{\sqrt{\Delta_{a} \Delta_{b} \Delta_{c}}}}{ \sqrt{\Theta(a,b,c)}}\,\, Vert[a,b,c].$$

\noindent {\bf Lemma.}  For the bracket evaluation at the root of unity $A = e^{i\pi/2r}$ the factor
$$f(a,b,c) = \frac{\sqrt{\sqrt{\Delta_{a} \Delta_{b} \Delta_{c}}}}{ \sqrt{\Theta(a,b,c)}}$$
is real, and can be taken to be a positive real number for $(a,b,c)$ admissible (i.e.  with $a + b + c \le 2r -4$).
\bigbreak

\noindent {\bf Proof.} See our basic reference \cite{KLSpie}.
\bigbreak

In \cite{KLSpie} we show how this modification of the vertex affects the non-zero term of the orthogonality of trivalent
vertices (compare with Figure 4). We refer to this as the ``modified bubble identity." The coefficient in the modified bubble identity is
$$\sqrt{ \frac{\Delta_{b}\Delta_{c}}{\Delta_{a}} } = (-1)^{(b+c-a)/2} \sqrt{\frac{[b+1][c+1]}{[a+1]}}$$ 
where $(a,b,c)$ form an admissible triple. In particular $b+c-a$ is even and hence this factor can be taken to be positive real.
\bigbreak

We rewrite the recoupling formula in this new basis and emphasize 
that the recoupling coefficients can be seen (for fixed external labels $a,b,c,d$) as a matrix transforming the horizontal ``double-$Y$" basis
to a vertically
disposed double-$Y$ basis. In Figures 10 and 11 we have shown the form of this transformation,using the matrix notation
$$M[a,b,c,d]_{ij}$$ for the modified recoupling coefficients. In Figure 11 we show an explicit formula for these matrix elements. The proof of this 
formula follows directly from trivalent--vertex orthogonality (See Figures 4 and 7.), and is given in \cite{KLSpie}. The result shown in
Figure 11 is the  following formula for the recoupling matrix elements.
$$M[a,b,c,d]_{ij} = ModTet
\left( \begin{array}{ccc}
a &  b & i \\
c & d & j \\
\end{array} \right)/\sqrt{\Delta_{a}\Delta_{b}\Delta_{c}\Delta_{d}}$$
where $\sqrt{\Delta_{a}\Delta_{b}\Delta_{c}\Delta_{d}}$ is short-hand for the product
$$\sqrt{ \frac{\Delta_{a}\Delta_{b}}{\Delta_{j}} }\sqrt{ \frac{\Delta_{c}\Delta_{d}}{\Delta_{j}} } \Delta_{j}$$ 
$$= (-1)^{(a+b-j)/2}(-1)^{(c+d-j)/2} (-1)^{j} \sqrt{ \frac{[a+1][b+1]}{[j+1]}}\sqrt{ \frac{[c+1][d+1]}{[j+1]}} [j+1]$$
$$ = (-1)^{(a+b+c+d)/2}\sqrt{[a+1][b+1][c+1][d+1]}$$
In this form, since
$(a,b,j)$ and $(c,d,j)$ are admissible triples, we see that this coeffient can be taken to be positive real, and its value is
independent of the choice of $i$ and $j.$
The matrix $M[a,b,c,d]$ is real-valued.
\bigbreak

\noindent It follows from Figure 10 (turn the diagrams by ninety degrees) that 
$$M[a,b,c,d]^{-1} = M[b,d,a,c].$$
Figure 10 implies the formula
$$M[a,b,c,d]^{T} = M[b,d,a,c].$$ It follows from this formula that 
$$M[a,b,c,d]^{T} = M[a,b,c,d]^{-1}.$$ {\it Hence $M[a,b,c,d]$ is an orthogonal, real-valued matrix.}

\begin{figure}
     \begin{center}
     \begin{tabular}{c}
     \includegraphics[height=1.5cm]{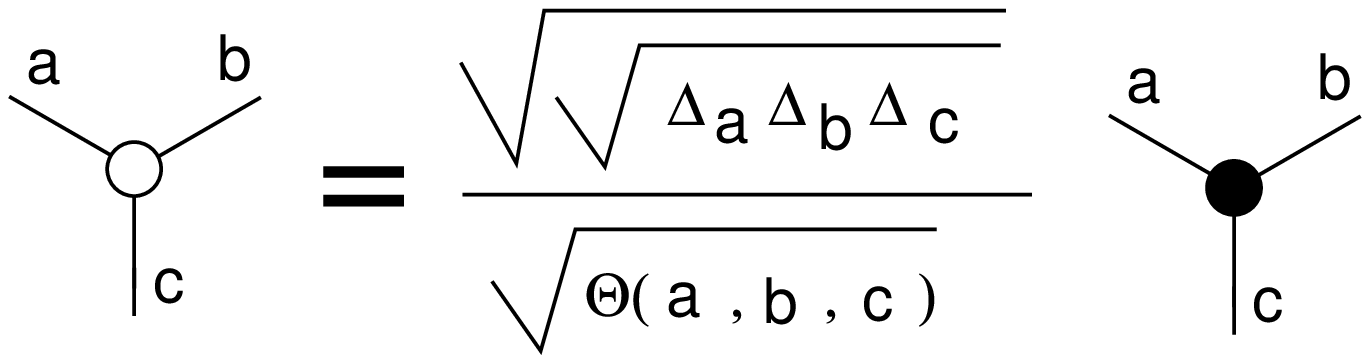}
     \end{tabular}
     \end{center}
     \caption{\bf Modified Three Vertex}
     \end{figure} 
     \bigbreak

\begin{figure}
     \begin{center}
     \begin{tabular}{c}
     \includegraphics[height=1.5cm]{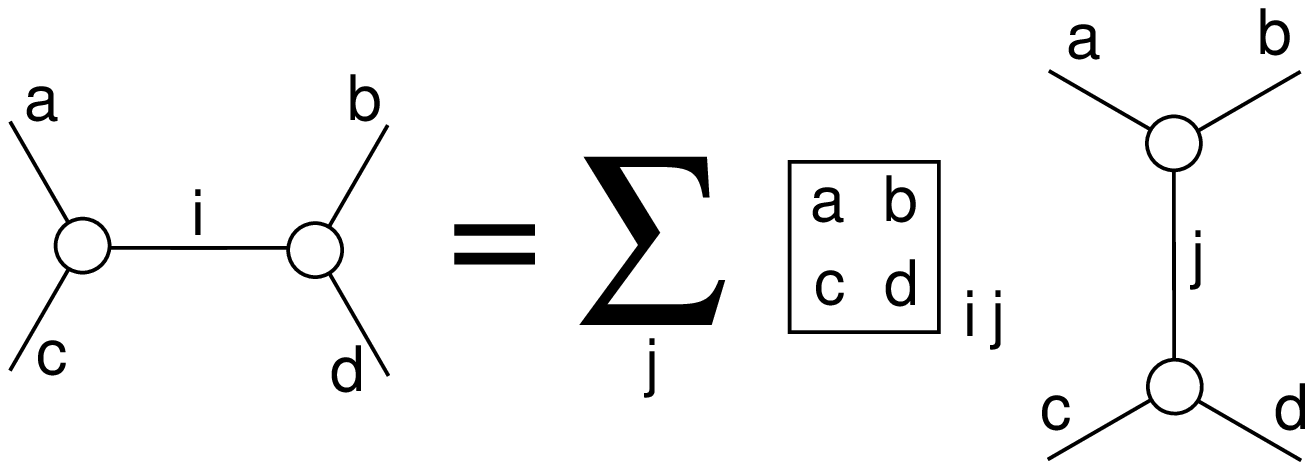}
     \end{tabular}
     \end{center}
     \caption{\bf Modified Recoupling Formula}
     \end{figure} 
     \bigbreak

\begin{figure}
     \begin{center}
     \begin{tabular}{c}
     \includegraphics[height=4cm]{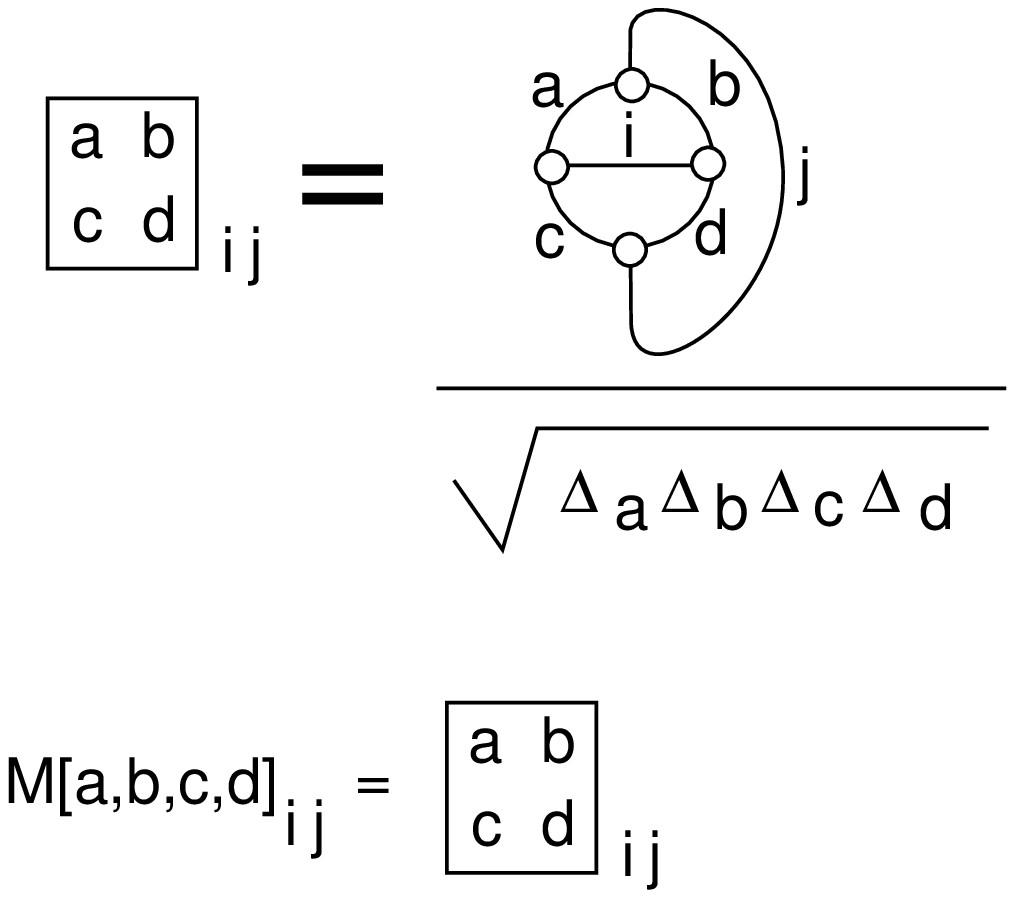}
     \end{tabular}
     \end{center}
     \caption{\bf Modified Recoupling Matrix}
     \end{figure} 
     \bigbreak

\noindent {\bf Theorem.} In the Temperley -- Lieb theory we obtain unitary (in fact real orthogonal) recoupling transformations when the bracket
variable $A$ has the form $A = e^{i\pi/2r}$. Thus we obtain families of unitary representations of the Artin braid group
from the recoupling  theory at these roots of unity. 
\bigbreak

\noindent {\bf Proof.} The proof is given by the discussion above and in \cite{KLSpie}. 
\bigbreak

\section{Explicit Form of the Braid Group Representations}
In order to have an explicit form for the representations of the braid group that we have constructed we return to the description of the vector spaces
in the introduction to this paper. Here we make this description of the vector spaces more precise as follows. We describe a vector space
$V[(a_{1}a_{2})a_{3}:a_{4}]$ depending upon a choice of three input and output spins where $(ab)$ denotes the possible outcome of two spin labels interacting at a
trivalent vertex as in Figure 3. In that figure we see that
$(ab)$ can represent
$c$ (the remaining leg of the vertex) and that there is a range of values possible for $c$ given by the constraints on $i$ $j$ and $k$ as shown in that
figure. Here we insist that the composite interaction $(a_{1}a_{2})a_{3}$ shall equal $a_{4}$ so that the vector space $V[(a_{1}a_{2})a_{3}:a_{4}]$ corresponds to
the left-hand tree shown in Figure 12. In that figure we indicate the recoupling mapping $F:V[(a_{1}a_{2})a_{3}:a_{4}] \longrightarrow V[a_{1}(a_{2}a_{3}):a_{4}].$
The matrix form of $F$ is composed from the recoupling matrix of Figure 11. In Figure 12 we have labeled $x = (a_{1}a_{2})$ corresponding to one of the basis
vectors in  $V[(a_{1}a_{2})a_{3}:a_{4}].$ Similarly, we have $y = (a_{2}a_{3})$ corresponding to one of the basis vectors in $V[(a_{1}a_{2})a_{3}:a_{4}].$
We let the corresponding vectors be denoted by $|x\rangle$ and $|y\rangle$ respectively. Then we can write
$$F|i\rangle = \Sigma_{j}F_{ji}|j\rangle$$ where $j$ ranges over the admissible labels for the interaction of $a_2$ and $a_3$.
\bigbreak

\begin{figure}
     \begin{center}
     \begin{tabular}{c}
     \includegraphics[height=2cm]{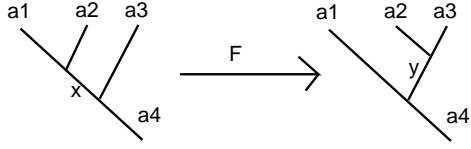}
     \end{tabular}
     \end{center}
     \caption{\bf Recoupling Map $F:V[(a_{1}a_{2})a_{3}:a_{4}] \longrightarrow V[a_{1}(a_{2}a_{3}):a_{4}]$}
     \end{figure} 
     \bigbreak

\noindent To see how the three strand braid group acts on  $V[(a_{1}a_{2})a_{3}:a_{4}],$ view Figure 13. If we let $s_1$ denote the generator of the three-stand
braid group $B_3$ that twists the first two strands and $s_2$ denote the generator that twists the second two strands, then we see that $s_1$ acts directly at
a trivalent vertex, giving the formula $$s_{1}|x\rangle = \lambda(a_1,a_2,x) |x\rangle$$ where $\lambda(a_1,a_2,x) = \lambda^{a_1,a_2}_{x}$ is the braiding
factor of Figure 8. On the other hand, we need to perform a recoupling in order to compute the action of $s_{2}.$ As shown in Figure 13, we have
$$s_{2}|i\rangle = \Sigma_{kj}F^{-1}_{kj}\lambda(a_3,a_4,j)F_{ji}|k\rangle.$$ This gives a complete description of the representation of the three-strand 
braid group on the vector space  $V[(a_{1}a_{2})a_{3}:a_{4}].$ Our next task is to generalize this to an abitrary ``left-associated" tree.
\bigbreak

\begin{figure}
     \begin{center}
     \begin{tabular}{c}
     \includegraphics[height=6cm]{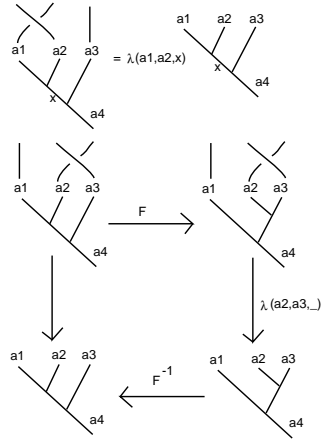}
     \end{tabular}
     \end{center}
     \caption{\bf Action of the Braid Group}
     \end{figure} 
     \bigbreak

\noindent We wish to consider larger left associated trees such as $V[((((a_{1}a_{2})a_{3})a_{4})a_{5}):a_{6}]$ To this purpose it is useful to 
declare that a fully left-associated product may be written without parentheses. Thus we have
$$a_{1}a_{2}a_{3}a_{4}a_{5} = ((((a_{1}a_{2})a_{3})a_{4})a_{5}) $$ and
$$a_{1}(a_{2}a_{3})a_{4}a_{5} = (((a_{1}(a_{2}a_{3}))a_{4})a_{5}).$$
Thus we have the recoupling transformation
$$F^{2}:V[a_{1}a_{2}a_{3}a_{4}a_{5}:a_{6}] \longrightarrow V[a_{1}(a_{2}a_{3})a_{4}a_{5}:a_{6}]$$
that will be used for the action of $s_{2}$ on the space $V[a_{1}a_{2}a_{3}a_{4}a_{5}:a_{6}].$ 
\bigbreak

In the general case we have the spaces $V[a_{1}a_{2} \cdots a_{n}:a_{n+1}]$ with basis elements 
$|x_{2}x_{3} \cdots x_{n-1}\rangle$ where
$(a_{1}a_{2})$ has $x_2$ as an outcome, $(x_{2}a_{3})$ has $x_3$ as an outcome, and so on until $(x_{n-1}a_{n})$ has $a_{n+1}$ as an outcome.
For articulating the braiding we need mappings
$$F^{i}: V[a_{1}a_{2} \cdots a_{n}:a_{n+1}] \longrightarrow V[a_{1}a_{2} \cdots a_{i-1}(a_{i}a_{i+1})a_{i+2} \cdots a_{n}:a_{n+1}].$$
The target space has the strands labeled $i$ and $i+1$ combined at a vertex so that the braiding for $s_i$ in the target space is local.
We also need a basis for $V[a_{1}a_{2} \cdots a_{i-1}(a_{i}a_{i+1})a_{i+2} \cdots a_{n}:a_{n+1}].$ This is given by the kets
$|y_{2}y_{3} \cdots y_{n-1}\rangle$ where 
$$(a_{1}a_{2}) = y_2$$
$$\cdots$$
$$(y_{i-2}a_{i-1}) = y_{i+1}$$
$$(a_{i}a_{i+1}) = y_{i}$$
$$(y_{i+1}a_{i+2}) = y_{i+2}$$
$$\cdots$$
$$(y_{n-2}a_{n-1}) = y_{n-1}$$
\bigbreak

\noindent We then have $$s_{i}|x_{2}x_{3} \cdots x_{n-1}\rangle = (F^{i})^{-1}\lambda(a_{i},a_{i+1})F^{i}|x_{2}x_{3} \cdots x_{n-1}\rangle.$$
Here it is understood that $$\lambda(a_{i},a_{i+1})|y_{2}y_{3} \cdots y_{n-1}\rangle = \lambda(a_{i},a_{i+1},y_i)|y_{2}y_{3} \cdots y_{n-1}\rangle,$$ where
$\lambda(a,b,c)$ is defined as explained above. Finally, using the recoupling matrix formalism of Figure 10, we have
$$F^{i}|x_{2}x_{3} \cdots x_{n-1}\rangle = \Sigma_{y}M[a_{i},a_{i+1},x_{i-1},x_{i+1}]_{y x_{i}}|x_{2}x_{3} \cdots x_{i-1}yx_{i+1}\cdots x_{n-1}\rangle.$$
This completes our description of the action of the braid group on these vector spaces.
\bigbreak

\subsection{The Fibonacci Model}
In the Fibonacci model \cite{Anyonic}, there is a single non-trivial recoupling matrix $F.$ 
$$F =
\left( \begin{array}{cc}
1/\Delta & 1/\sqrt{\Delta} \\
1/\sqrt{\Delta} & -1/\Delta \\
\end{array} \right) =
\left( \begin{array}{cc}
\tau & \sqrt{\tau} \\
\sqrt{\tau} & -\tau \\
\end{array} \right)$$
where $\Delta = \frac{1 + \sqrt{5}}{2}$ is the golden ratio and $\tau = 1/\Delta$.
The local braiding matrix is given by the formula
below with $A = e^{3\pi i/5}.$
$$R = 
\left( \begin{array}{cc}
A^{8} & 0 \\
0 & -A^{4} \\
\end{array} \right)=
\left( \begin{array}{cc}
e^{4\pi i/5} & 0 \\
0 & -e^{2\pi i/5} \\
\end{array} \right).$$
\bigbreak

The simplest example of a braid group representation arising from this theory is the representation of the three strand braid group generated by
$s_{1}= R$ and $s_{2} = FRF$ (Remember that $F=F^{T} = F^{-1}.$). The matrices $s_{1}$ and $s_{2}$ are both unitary, and they generate a dense subset of
$U(2),$ supplying the local unitary transformations needed for quantum computing. 
\bigbreak

\begin{figure}
     \begin{center}
     \begin{tabular}{c}
     \includegraphics[height=4cm]{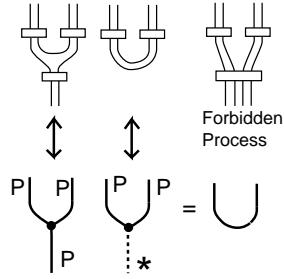}
     \end{tabular}
     \end{center}
     \caption{\bf Fibonacci Vertices}
     \end{figure} 
     \bigbreak

In the Fibonacci model there are two labels, as we described in the introduction (see Figure 14): $P$ and $*$. $P$ can interact with itself to produce
either
$P$ or $*,$ while $*$ acts as an identity element. That is, $*$ interacts with $P$ to produce only $P,$ and $*$ interacts with $*$ to produce $*.$ Let
$$V[n] = V[a_{1}a_{2} \cdots a_{n}:a_{n+1}] = V(P P P \cdots P:P).$$ The space $V[n]$ has basis vectors
$|x_{2}x_{3}\cdots x_{n-1} \rangle$ where $\{x_{2},x_{3},\cdots x_{n-1} \}$ runs over all sequences of $P$'s and $*$'s without consecutive $*$'s.
The dimension of $V[n]$ is $f_{n},$ the $n$-th Fibonacci number: 
$f_{1}=1, f_{2}=1, f_{3} = 2, f_{4}= 3, f_{5} = 5, f_{6}=8, \cdots$ and 
$f_{n+1} = f_{n} + f_{n-1}.$ 
\bigbreak

In terms of the matrix $R$, we have and $\lambda(*) = A^{8}$ and $\lambda(P) = - A^{4}.$ The representation of the the braid group $B_{n}$ on $V[n]$ is
given by the formulas below (with $x_{0} = x_{n} = P$ and $i = 1,2,\cdots n-1$ and the matrix indices for $F$ are $*$ and $P$ corresponding to 
$0$ and $1$ respectively). We use the matrix $N = FRF$ below.
$$s_{1}|x_{2}x_{3} \cdots x_{n-1} \rangle = \lambda(x_{2})|x_{2}x_{3}\cdots x_{n-1} \rangle,$$ and for $i \ge 2:$
$$s_{i}|x_{2}x_{3} \cdots x_{n-1} \rangle = \lambda(x_{i})|x_{2}x_{3}\cdots x_{n-1} \rangle$$ if $x_{i-1} \ne P$ or $x_{i+1} \ne P.$
$$s_{i}|x_{2}x_{3} \cdots x_{n-1} \rangle = \Sigma_{\alpha = *,P} N_{\alpha, x_{i}}|x_{2}x_{3} \cdots x_{i-1}\, \alpha \,\, x_{i+1} \cdots  x_{n-1} \rangle$$ 
if $x_{i-1} = x_{i+1} = P.$ 

\noindent These formulas make it possible to do full-scale computer experiments with the Fibonacci model and the generalizations of it that we have discussed.
We will pursue this course in a subsequent paper. This model is universal for quantum computation.
\bigbreak

\section*{ACKNOWLEDGMENTS}  
The first author thanks the National Science Foundation for support of this research under NSF Grant DMS-0245588.
\bigbreak

 \end{document}